\documentclass[nofootinbib, superscriptaddress]{article}
\RequirePackage[left = 2.5cm, right = 2.5cm, top = 2.5cm, bottom = 3.5cm]{geometry}
\pdfoutput=1

\usepackage[utf8]{inputenc}
\usepackage{graphicx}
\usepackage{subfigure}
\usepackage{cite} 
\usepackage{placeins}
\usepackage{slashed}
\usepackage{verbatim}
\usepackage{amssymb}
\usepackage{amsmath}
\usepackage{makecell}
\usepackage[title]{appendix}
\usepackage{etoolbox}
\patchcmd{\appendices}{\quad}{: }{}{}
\usepackage{hyperref}
\usepackage{caption}
\usepackage{authblk}
\usepackage{color}
\usepackage{xcolor}
\usepackage{hyperref}
\hypersetup{
	colorlinks = true,
	linkcolor  = red,
	citecolor = blue,
	filecolor = cyan,
	menucolor = red,
	urlcolor = magenta
	}

\newcommand{\nicpb}{\emph{National Institute of Chemical Physics and Biophysics, Akadeemia tee 23, 12618 Tallinn, Estonia}}

\newcommand{\ut}{\emph{Institute of Physics, University of Tartu; Ravila 14c, 50411 Tartu, Estonia.}}

\newcommand{\td}{\mathrm{d}}

\begin{document}

\title{A new mechanism for dark energy: the adaptive screening}

\author[1]{Andi Hektor}
\author[2]{Luca Marzola}
\author[1,2]{Martti Raidal}
\author[2]{Hardi Veerm\"ae}

\affil[1]{\nicpb}
\affil[2]{\ut}

\date{\today}

\maketitle

\begin{abstract}
We describe how known matter effects within a well-motivated particle physics framework can explain the dark energy component of the Universe. By considering a cold gas of particles which interact via a vector mediator, we show that there exists a regime where the gas reproduces the dynamics of dark energy. In this regime the screening mass of the mediator is proportional to the number density of the gas, hence we refer to this phenomenon as ``the adaptive screening mechanism''. As an example, we argue that such screening mass can result from strong localization of the vector mediators. The proposed dark energy mechanism could be experimentally verified through cosmological observations by the Euclid experiment, as well as by studying properties of dark photons and sterile neutrinos. 
\end{abstract}

\section{Introduction}

Cosmological observations have progressively established that about 68\% of the total energy density of the Universe is in the form of a mysterious agent, the dark energy. This evasive component is usually modelled in the equation of state $p= w \rho$, which relates the components' pressure $p$ to its energy density $\rho$ through the parameter $w$. The dedicated experiments currently measure the latter at 95\% confidence level in the interval $w = -1.13^{+0.13}_{-0.14}$ and set the characteristic energy scale at $\Lambda_{0}^{1/4} \simeq 10^{-3}\text{ eV}$~\cite{Garnavich:1997nb, Riess:1998cb, Perlmutter:1998np, Boughn:2003yz, Scranton:2003in, Fosalba:2003ge, Tegmark:2003ud, Astier:2005qq, Wright:2008ib, Komatsu:2010fb, Ade:2013ktc}. The same measurements allow for the suggestive interpretation of dark energy as a small but non-vanishing pure vacuum energy: the famous cosmological constant, for which $w=-1$ must hold at all times. However, within the standard model of particle physics, zero-modes of the fields, the quark-gluon condensate at the QCD phase transition or the Higgs bosons vacuum expectation value should all induce sizeable contributions into the former. 
As none of these contributions has been observed in Nature, it seems plausible that some unknown mechanism is at work to set a vanishing vacuum energy, enforcing by net the cancellations of the mentioned quantum effects. However ad-hoc such idea might seem, the corresponding point of view has already been adopted in contemporary physics. An example is brought by quintessence theories  \cite{Peebles:1987ek, Ratra:1987rm}, where a nearly-massless slow-rolling scalar field accounts for the measured dark energy density on top of a vanishing vacuum contribution. In the present paper we also assume that vacuum has a null energy density, with the purpose of introducing a new, dynamical mechanism to explain the origin of dark energy. 

Here we consider a natural setup for the dark sector, where the particles of a cold and diluted gas are coupled to a light vector field, the dark photon~\cite{Essig:2013lka}. In this framework we show that whenever the adaptive screening is active, i.e. whenever the mediator mass is dominated by an effective screening mass proportional to the number density of the gas particles, then the gas may enter a regime in which the equation of state parameter is $w \approx -1$ and therefore act as a dark energy.

As an example, we choose to put forward the strong, or Anderson, localization of the vector mediator as a possible realization of the adaptive screening mechanism. The localization of electrons was first discussed by Anderson in the context of the metal-to-insulator transition in the presence of impurities~\cite{Anderson:1958vr}. It was later realized that  localization transitions are general wave phenomena that occur in random media with dimensionality of two or larger, roughly speaking due to the destructive interference between the waves scattered from the randomly distributed scattering centers. The latter, in our case, are represented by the gas particles. To date localization effects have been observed in a variety of systems ranging from electromagnetic~\cite{Wiersma:2004,Chabanov:2004,Storzer:2006} to acoustic waves~\cite{Strybulevych:2008}. 

In the following, after presenting the basis of the new mechanism, we assume that the gas is composed of a single species of light fermions $\psi$ and show how the measurements of dark energy density cast a stringent bound on the mass of the fermions itself, $\Lambda_{0}^{1/4} \propto m_{\psi}\approx 10^{-3}$~eV. The resulting mass scale is typical to neutrino physics and, interestingly, an extra ``sterile'' neutrino with a squared mass splitting $\Delta m^2_{1\psi}\approx 10^{-5}$~eV$^2$ has previously been proposed in literature as a solution to the upturn problem of the solar neutrino flux~\cite{deHolanda:2010am,Mirizzi:2013kva}. Our framework also predicts the existence of dark photons which will be investigated in dedicated experiments~\cite{Essig:2013lka}. Furthermore, upcoming cosmological observations by the Euclid experiment~\cite{Amendola:2012ys} may discriminate between our mechanism and a pure cosmological constant.

\section{The mechanism}

To illustrate how the adaptive screening mechanism works, let us consider a diluted gas of non-relativistic particles coupled via a long range repulsive force\footnote{A repulsive potential ensures that the energy density of the system is always positive and guarantees the local stability of gas.}. The latter is characterised by the potential $U_{\rm int}(r_{ij})$ between each couple of particles labeled by $i$ and $j$. If the total number of particles is conserved, the equation of state parameter can be expressed in terms of the total energy density $\rho$ and the number density of the gas particles $n$ as in\footnote{Two possible derivations are presented in Appendix \ref{app_A}.}
\begin{equation}\label{eq:eos}
 	w := \frac{p}{\rho} = \frac{\partial \log \rho}{\partial \log n}-1. 
\end{equation}
We also require that the vacuum, defined as the state with $n=0$, has a vanishing energy density, $\rho(n=0) = 0$. As a consequence, every dark energy regime, for which $w = -1$, has necessarily to be established dynamically.

The total energy density of the non-relativistic gas comprises two  contributions, given respectively by the rest mass $M$ of particles and by their interaction energy
\begin{equation}\label{eq:rho_gas}
	\rho(n) = M n + \rho_{\rm int}(n),
	\qquad
	\rho_{\rm int}(n):= U_{\rm int}/V.
\end{equation}
Plugging the above equations into Eq.~\eqref{eq:eos} gives
\begin{align}\label{eq:general_w}
	w = \frac{M n}{M n + \rho_{\rm int}}\left(1 + \frac{1}{M}\frac{\partial \rho_{\rm int}}{\partial n}\right) - 1.
\end{align}
In a regime where $M n \ll \rho_{\rm int}$ and the dependence of $\rho_{\rm int}$ on $n$ is negligible, the gas is characterised by $w \approx -1$ and reproduces the desired dark energy behaviour. This observation is at the basis of the adaptive screening mechanism.

Assuming a repulsive potential and an approximately uniform number density distribution, the total potential energy density in Eq.~\eqref{eq:rho_gas} is given by \cite{Stiele:2010xz}
\begin{align}\label{eq:potenergy}
	\rho_{\rm int} \approx \frac{g^{2} n^{2}}{2m^{2}},
\end{align}
where $g$ is the coupling between the gas constituents and the interaction mediators having a mass $m$.  In the diluted cold gas approximation, once localization is active, the former comprises two contributions,
\begin{align}\label{eq:effmass}
 	m(n) = m_0 + \sigma n.
\end{align}
Here $m_0$ is the invariant mass of the mediator while the term $\sigma n$ represents the effect of the non-perturbative screening. The presence of the second contribution implements the adaptive screening mechanism. On dimensional grounds we expect the effective elastic scattering cross-section of the mediator on a gas particle to be $\sigma = \kappa \, g^4/M^2$, where typically $\mathcal{O}(10^{-2})\lesssim\kappa\lesssim \mathcal{O}(10^{-1})$.

With the above results we can now study the behaviour of $\rho(n)$ and $w$ in different regimes. Plugging Eq.~\eqref{eq:potenergy} and \eqref{eq:effmass} into Eq.~\eqref{eq:rho_gas} and \eqref{eq:general_w} yields
\begin{align}
	\rho &=  M n +  \Lambda\left(1 + n_2/n\right)^{-2},\label{eq:rho}
	\\
	w &= \frac{n_2 - n}{n_2 + n}\left(1 + n/n_1 (1 + n_2/n)^{2}\right)^{-1},\label{eq:w_n}
\end{align}
where we defined the characteristic densities
\begin{gather}
	\Lambda := \frac{g^2}{2\sigma^{2}} = \frac{M^{4}}{2\kappa^{2} g^{6}},	\qquad 
	n_1 = \frac{\Lambda}{M}, \qquad 
	n_2 = \frac{m_0}{\sigma}, \qquad 
	n_3 = \frac{n_2^{2}}{n_1}.\label{eq:char_densities}
\end{gather}

\begin{figure}[h]
 \begin{minipage}[c]{.49\textwidth}
	 \centering
 \includegraphics[width=.95\textwidth]{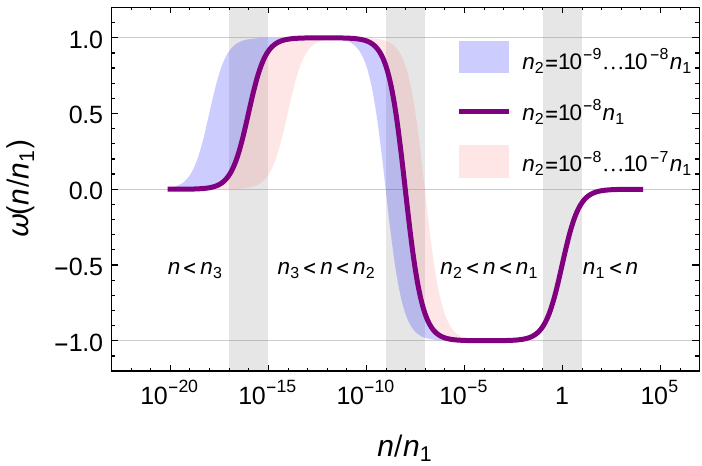}
 \captionof{figure}{The equation of state parameter $w$ as a function of $n/n_1 \propto (1+z)^3$ for different values of the ratio $n_1/n_2$, see Eq.~\eqref{eq:w_n}. The blue and pink bands show how the function $w(n)$ depends on the ratio $n_1/n_2$. Notice that the dark energy regime $w = -1$ can be extended arbitrarily by choosing a suitable ratio of $n_1/n_2 \propto 1/m_0$. The gray areas identify fast transition regimes where the number density $n$ changes by about two orders of magnitude.}
 \label{fig:w}
 \end{minipage}
 \hspace{.2 cm}
 \begin{minipage}[c]{.40\textwidth}
	 \centering
	 \begin{tabular}{cccp{.45\textwidth}}
	 \hline\hline
	 {Number density} & {$w$} & {$\rho$} & {Description}\\ \hline
	 $n\ll n_3$ & 0 & $M n$ & {\it Low density dust}, the interaction is completely screened \\
	 $n_3 \ll n \ll n_2$ & 1 & $\frac{g^{2}n^{2}}{2 m_0^{2}}$ & {\textit{Stiff fluid}, the screening is only due to $m_0$} \\
	 $n_2 \ll n \ll n_1$ & -1 & $\Lambda$ & {\textit{Dark energy}, the adaptive screening due to $\sigma n$ dominates} \\
	 $n_1 \ll n $ & 0 & $M n$ & {\textit{High density dust}, the rest mass $M$ dominates} \\
	 \hline\hline
	 \end{tabular}
	 \captionof{table}{The equation of state parameter $w$ as a function of the gas constituents number density $n$. The characteristic densities $n_{1,2,3}$ and $\Lambda$ are defined in Eq.~\eqref{eq:char_densities}.}
	 \label{tab:density_regions}
	 \end{minipage}
\end{figure}

As made clear by Fig.~\ref{fig:w} and Tab.~\ref{tab:density_regions}, the above definitions identify four regimes distinguished by different values of the equation of state parameter of the gas. In the first regime $n\gg n_1$, the rest mass of the gas constituents dominates the energy density of the system, which therefore behaves like dust: $w=0$. As the number density drops (due to the expansion of the Universe) to $n_2\ll n \ll n_1$, the localization effects prevail and the adaptive screening mechanism ensures an approximately constant energy density. In this regime we recover the dark energy behaviour, $w \approx -1$, while the energy density itself converges to $\Lambda$. Such dynamics is maintained until $n_3\ll n \ll n_2$, where the rest mass of the vector mediator is setting the screening length of the potential. In this stage the interaction energy of the system is quickly depleted owing to $w\approx 1$. Below the threshold $n= n_3$ the rest mass contribution of the gas constituents tops the interaction one and $w=0$ is established once again. The last two regimes are also predicted  
in studies of self-interacting dark matter~\cite{Stiele:2010xz}, confirming our results in the absence of the adaptive screening effect.

In the above derivation we assumed $n_2 \lesssim 10^{-4} \, n_1$, which translates in the following bound on the mediator mass $m_0$,
\begin{equation}\label{eq:smallxi}
 m_0 \lesssim 10^{-4} \frac{M}{2 g^2}.
\end{equation}

Furthermore, when a gas of non relativistic fermions is considered in our setup, setting the dark energy density to the present value $\Lambda \approx (10^{-3}\text{ eV})^4$ and assuming $g = \mathcal{O}(1)$ yield $M \approx 10^{-3}$~eV and the upper bound $m_0 \lesssim 10^{-7}$~eV. The proposed mechanism could then be tested through the predicted particle content, in particular via its possible implications in neutrino physics which we discuss in a dedicated Section.

When the background of an expanding Universe is considered, the requirement that the interaction mediators remain efficiently coupled to gas imposes a further constraint on the relevant interaction rate $\Gamma_{int} = n\sigma(n)$,
\begin{equation}\label{eq:cosmn}
	\frac{\Gamma_{\rm int}}{H}\gtrsim 1,
\end{equation}
where $H$ is the Hubble parameter. Effectively, the above equation imposes that the localization length, $\sim 1/\sigma n$, never exceeds the Hubble distance.  Beside that, the relation $n_* \sigma(n_*) = H$ implicitly defines the limit number density $n_*$ such that for $n<n_*$ the interactions in the gas are negligible and therefore $w=0$. Hence, every realistic implementation of our mechanism must respect the condition $n > n_*$ while the dark energy regime is maintained.

\section{Cosmological implications}

\begin{figure}[h]
	\centering
 \includegraphics[width=.70\textwidth]{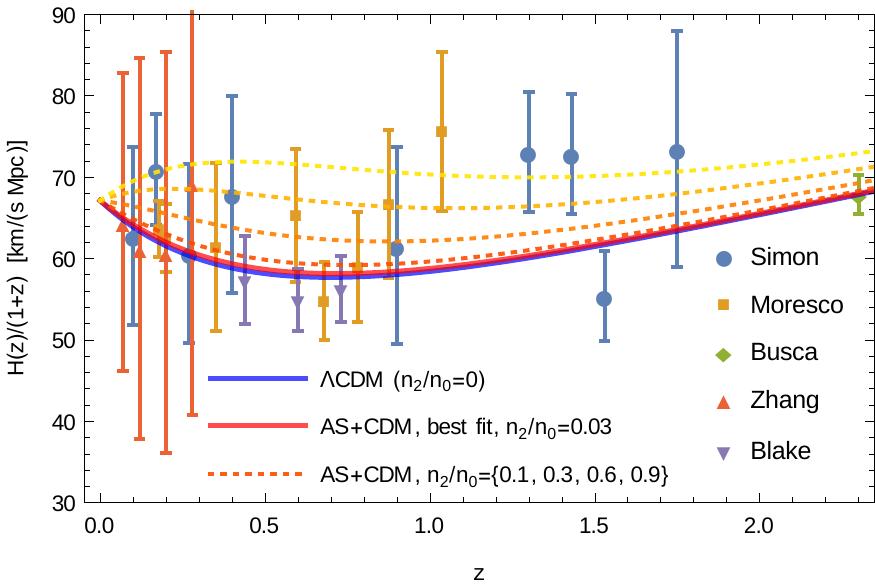}
 \caption{The Hubble parameter plotted against the redshift. The presented data and the corresponding 68.6\% confidence interval is taken from \cite{Farooq:2013hq}. The dashed lines represent our model for different values of the ratio $n_2/n_0$ while the solid blue line corresponds to the $\Lambda$CDM model, that is a limiting case of our model ($n_2 = 0$). The solid red line shows the best fit of our model, obtained for $n_2 = 0.0(3)$, $\Omega_M = 0.29(5)$, $H_0 = 68(7) \, \mathrm{km \, s^{-1}Mpc^{-1}}$.}
 \label{fig:h}
\vspace{-\baselineskip}
\end{figure}

In our setup, the total number of gas particles in a comoving volume is conserved. By setting $a_0 := a(z=0) = 1$, the number density of our gas constituents scales as $n = n_0 a^{-3} \propto (1 + z)^{3}$ and serves as an effective measure of the cosmological time. The equation of state parameter then evolves through the regimes described in Fig.~\ref{fig:w} and Tab.~\ref{tab:density_regions} as the gas is progressively diluted. How our mechanism affects the evolution of the Universe is quantified by the first Friedmann equation
\begin{equation}
	\left(\frac{H}{H_{0}}\right)^{2} = \frac{\rho}{\rho^c_{0}}
\end{equation}
where $H := \dot{a}/a$ is the Hubble parameter, $\rho^{c}_{0} := 3 H_{0}^{2}/8\pi G$ is the critical energy density and $G$ is the gravitational constant. Inserting \eqref{eq:rho} into the above equation, as well as neglecting the contributions of radiation and curvature, results in
\begin{align}\label{I_Friedmann_eq}
	\frac{H^{2}}{H_{0}^{2}} 
	=	\Omega_{M} a^{-3} 
	+ \Omega_{\Lambda} \left(\frac{n_0 + n_2}{n_0 + n_2 a^{3}}\right)^{2}
	,
\end{align}
where $n_0$ is the current number density of the gas constituents. The density parameter 
\begin{align}\label{Omega_L}
	\Omega_{\Lambda} := \frac{\Lambda/\rho^c_0 }{1 + n_2/n_0}
\end{align}
accounts for the potential energy density in \eqref{eq:rho} and results in the required dark energy behaviour provided that $n_2 \ll n_0$. The matter contribution $\Omega_{M} := \Omega_{\psi} + \Omega_{DM} + \Omega_{b}$ comprises respectively the rest mass contribution of the proposed cold gas particles,  
\begin{align}\label{Omega_psi}
	\Omega_{\psi} 
	:= \frac{M n_0}{\rho_c} 
	\approx \frac{n_0}{n_1} \Omega_{\Lambda} 
\end{align}
the Dark Matter one, $\Omega_{DM}$, and the baryonic matter density parameter $\Omega_{b}$. Presently $\Omega_{\Lambda} \approx 68\%$ and $\Omega_{M}\approx 31\%$\cite{Ade:2013ktc} and the resulting evolution of the Hubble parameter, Eq.~\eqref{I_Friedmann_eq}, is shown in Fig.~\ref{fig:h} for different values of the ratio $n_2/n_0$.  

As made clear by the first Friedmann equation, Eq.~\eqref{I_Friedmann_eq},  $\Omega_{\psi}$ constitutes by net a fraction of the cold dark matter content of the Universe. However, given the characteristic mass scale $M \approx 10^{-3}$ eV, because of phase space constraints and the Pauli blocking argument \cite{Hannestad:2006zg}, our cold gas can only account for a sub-dominant fraction of the detected dark matter abundance. Interestingly, a self-interacting dark matter component is motivated by the missing satellite, core-vs-cusp and too-big-to-fail problems~\cite{Tulin:2013teo}, and could have implication on the dynamics of structure formation.

As shown in Fig.~\ref{fig:w} and Eq. \eqref{Omega_psi}, requiring the dark energy regime at present time implies that the number density $n$ exceeded $n_1$ in the past. The system entered the current accelerating regime after the number density has been diluted by the expansion of the Universe. A further characteristic of our framework is that the current dark energy regime will be abandoned in the future. For $n \lesssim n_2$, the gas enters a stiff fluid regime and its energy density is diluted as $\rho \propto a^{-6}$ with the expansion of the Universe. Early phases of such transition could then be detected by the forthcoming Euclid experiment as a downturn in the dark energy component. As long as this transition takes place far enough in the future, cosmological measurements cannot distinguish between the propose scheme and the $\Lambda$CDM model. The current observations \cite{Farooq:2013hq} impose $n_2 < 0.3 \, n_0$ at 68.6\% confidence level.

\section{Adaptive screening from localization}

We focus now on the behavior of the gas in a cold and diluted limit. With the gas particles acting as stationary scattering centers, the virtual mediators are localized consequently to the effect of the randomness in the environment. This provides a way to implement the adaptive screening mechanism in our system. Despite Anderson localization having being proposed more than fifty years ago~\cite{Anderson:1958vr} and having being observed in a large variety of systems~\cite{Wiersma:2004,Chabanov:2004,Storzer:2006,Strybulevych:2008}, to date the localization transition has been proven in three spatial dimensions only, by using numerical methods. Therefore, we present the necessary conditions for localisation to happen and, based on the analogy with systems in which this phenomenon indeed occurs, we argue that localization is also a characteristic feature of our setup.

In three spatial dimensions, the Ioffe-Regel condition $\ell \lesssim \lambda$ provides an approximate criterion for distinguishing between the non-localized and localized regimes. Here $\ell = \left(\sigma n\right)^{-1}$ is the mean free path of a wave and $\lambda$ its wave length~\cite{Ioffe:1960aa}. If the Ioffe-Regel condition is satisfied, the propagator over a distance $x$ is then suppressed as $\exp({-\xi x})$ because of localization. Estimates of the localization length $\xi$ are given in literature, yielding $\xi \approx \ell$~\cite{Aizenman:1993st, Bershadsky:2003se}. In the localized regime, on top of the proper mass contribution in Eq.~\eqref{eq:effmass}, the vector mediator consequently receives an effective screening mass $\xi^{-1} \approx \sigma n$. The range of the interactions mediated by the vector particle is then constantly adapting to the changes in the density of the fermion gas, in a way that the interaction energy density of the gas is held approximately constant. The adaptive screening mechanism is therefore active and, for Eq.~\eqref{eq:general_w}, the system reproduces the desired dark energy dynamics.  \\

\section{A link to neutrino physics}

As pointed out before, imposing the current dark energy constraints results in a light mass scale $M \approx 10^{-3}$ eV. This suggests the existence of a new sterile neutrino species and, certainly, it is suggestive that a particle with compatible properties is currently being investigated in relation to the solar neutrinos upturn problem~\cite{deHolanda:2010am}. The SNO, Super-Kamiokande and Borexino experiments have not detected the upturn in the spectrum of solar neutrino events expected at energies of few MeV. A possible solution calls for a very light sterile neutrino species $\psi$ characterized by the mass square difference $\Delta m^2_{1 \psi}=m_\psi^2-m_1^2~(0.7-2) \times 10^{-5}$~eV$^2$ with respect to the lightest active neutrino mass $m_1$~\cite{deHolanda:2003tx}. Once a small mixing with the corresponding neutrino is provided, $\sin^2(2\theta_{1\psi})\sim10^{-3}$, the sterile neutrino reduces the survival probability of the electron component that otherwise causes the expected upturn. Interestingly, if $m_\psi \gg m_1$, the above solution points to the same mass scale $m_\psi \approx 10^{-3}$~eV that dark energy measurements indicate in our model, giving rise to a new interplay between neutrino physics and late time cosmology.

A proper investigation of this link is certainly model dependent. Proposing a specific framework is beyond the scope of the paper and therefore we will only discuss the general constraints that affect this kind of scenario. Cosmology plays a key role in testing the above connection, for instance through the bounds on the absolute (active) neutrino mass scale which effectively constrain $m_1$~\cite{Ade:2013zuv}. Alternatively, big bang nucleosynthesis and analyses of the cosmic microwave background radiation might test the proposed scenario by limiting the number of thermalised relativistic degrees of freedom at the corresponding epochs. Whether it seems unlikely that the required neutrinos yield a sizeable contribution to this quantity owing to the small mixing and large asymmetry required \cite{Archidiacono:2014nda,deHolanda:2010am}, the massive dark photon which mediates their interactions can effectively constrain the scenario \cite{Essig:2013lka}.
On the particle physics side, the ``3+1'' neutrino scheme proposed to solve the upturn problem is consistent with the latest global analysis of the sector, \cite{Mirizzi:2013kva,Esmaili:2013yea} and is currently subject to a dedicated investigation through reactor experiments \cite{Palazzo:2013bsa,Bakhti:2013ora}.  

\section{Conclusions}

In this paper we have shown that known matter effects can give rise to the dark energy component of the Universe. Our setup consists of a cold gas of particles interacting via a vector mediator, the dark photon, characterised by an environmentally induced mass that is proportional to the number density of the gas constituents. This implements ``the adaptive screening mechanism''  and allows our system to mimic the dynamics of dark energy through an approximately constant energy density that ensures $w \approx -1$. As an example, we put forward the Anderson localization as a possible realization of the adaptive screening mechanism. 
The proposed dark energy mechanism could be tested through cosmological observations by the Euclid satellite, through the detection of a downturn in the dark energy contribution to the energy budget of the Universe. 

A characteristic feature of our scenario is the presence of massive dark photons, currently subject of an intense experimental search. Once a fermion gas is considered, the proposed framework predicts a relation between the mass of the gas particles and the scale of dark energy, $\Lambda^{1/4} \approx \, M g^{-3/2}$ (Eq.~\eqref{eq:char_densities}). It is suggestive that current measurements of $\Lambda$ recover the mass scale $M\approx 10^{-3}$~eV already proposed within neutrino physics in connection to the upturn problem of the solar neutrino flux.

\section*{Acknowledgements}

This work was supported by grants MJD387, MTT60, IUT23-6, CERN+, and by EU through the ERDF CoE program. 

\begin{appendices}
\section{The equation of state}
\label{app_A}

Equation \eqref{eq:eos} follows from the conservation of energy and particle number. We present below two different derivation of this equation, based respectively on thermodynamics and on the continuity equation in an expanding geometry.
\begin{itemize}
	\item According to the first law of thermodynamics, in a isolated system $\td U + p \, \td V = \td Q$. Given the occupied volume $V$ and internal energy $U$, the pressure can be calculated as
$
	p = -\partial U/\partial V.
$
Identifying the internal energy with the total energy density $\rho := U/V$ then yields 
$
	p = - \rho - \td \rho/\td \ln V.
$
If the total number of particles $N = n V$ is conserved, $\td \ln V =  - \td \ln n$ and \eqref{eq:eos} follows.
	\item The continuity equation in the Friedmann–Lema\^{i}tre–Robertson–Walker Universe reads 
$
	\dot{\rho} = -3H(\rho + p).
$
This can be equivalently expressed as 
$
	\td\ln\rho = (\rho + p) \td\ln(a^{-3}).
$
If the particle number in a comoving volume is conserved, then the number density scales as $n \propto  a^{-3}$ so Eq. \eqref{eq:eos} follows.
\end{itemize}
\end{appendices}
%

\end{document}